\documentstyle[aps,prl,multicol,floats,epsf]{revtex}

\begin{document}
\draft
\wideabs{

\title{Low-temperature electronic heat transport 
in La$_{2-x}$Sr$_x$CuO$_4$ single crystals: \\
Unusual low-energy physics in the normal and superconducting states}

\author{J. Takeya, Yoichi Ando, Seiki Komiya, and X. F. Sun}
\address{Central Research Institute of Electric Power
Industry, Komae, Tokyo 201-8511, Japan}

\date{\today}
\maketitle

\begin{abstract}
The thermal conductivity $\kappa$ is measured 
in a series of La$_{2-x}$Sr$_x$CuO$_4$ ($0 \leq x \leq 0.22$) 
single crystals down to 90 mK to elucidate the evolution of the 
residual electronic thermal conductivity $\kappa_{\rm res}$, 
which probes the extended quasiparticle states in the $d$-wave gap.
We found that $\kappa_{\rm res}/T$ grows smoothly, except for a 1/8 
anomaly, above $x \simeq 0.05$ and shows no discontinuity at optimum
doping, indicating that the behavior of $\kappa_{\rm res}/T$ is not
governed by the metal-insulator crossover in the normal state;
as a result, $\kappa_{\rm res}/T$ is much larger than what the
normal-state resistivity would suggest in the underdoped region, 
which highlights the peculiarities in the low-energy physics in the 
cuprates.
\end{abstract}


}
\narrowtext

It is now generally perceived that the superconducting state of the
high-$T_c$ cuprates can be more or less conventionally described by
a BCS-like condensate with a $d$-wave symmetry and well-defined
Fermi-liquid-like quasiparticle (QP) excitations from it,
though the mechanism for the occurrence of the superconductivity
is expected to be highly unconventional.
Specifically, the $d$-wave phenomenology of the superconducting
state predicts such phenomena as the ``Volovik effect"
\cite{Volovik} and
the ``universal" heat conduction \cite{Lee},
both of which have been
confirmed by experiments; namely, in optimally-doped
YBa$_2$Cu$_3$O$_{7-\delta}$ (Y-123), the electronic specific heat has
been shown to increase with the magnetic field $H$ as $\sqrt{H}$
\cite{Moler}
and the electronic thermal conductivity for $T \rightarrow 0$ has been
shown to be independent of impurity concentration \cite{Taillefer}.
These effects are essentially caused by the QPs induced near the 
nodes of the $d$-wave gap either by vortices or by impurities.
In particular, impurities are believed to induce an ``impurity band"
at the Fermi energy $E_F$ in the $d$-wave gap, and the extended QPs
in this band are considered to form an ordinary Fermi liquid;
in this sense, one can say that the low-energy physics of the
cuprates in the superconducting state is governed by a Fermi
liquid \cite{Chiao}, while that in the normal state appears to be
governed by a non-Fermi liquid \cite{AndoBi2201}.

Recently, there appeared several studies which suggest that the
rather simple picture described above may not be the whole story.
It was theoretically argued, for example, that the impurity band may
actually develop a gap-like feature at very low energies near $E_F$
\cite{Senthil1999,Atkinson};
also, the quantum interference effect may lead to a localization of
the extended QP states \cite{Senthil1998}.
Experimentally, it was recently reported that the
QP contribution to the thermal conductivity is absent at low temperatures 
in an underdoped cuprate YBa$_2$Cu$_4$O$_8$ (Y-124) \cite{Hussey},
which indicates that there is no extended QP at low temperatures in 
this system; this behavior appears to be a result of some kind of 
localization of the QPs and thus is somewhat reminiscent of the 
unusual charge localization found in the normal state of the underdoped
cuprates under high magnetic fields \cite{Ando,Boebinger,Fournier,Ono}.
If the disappearance of the extended QPs in the superconducting state
and the ``insulating" normal state \cite{Boebinger} 
are actually related, that might mean
that the same mechanism dictates the charge localization in both the
normal and the superconducting states.
This issue becomes even more intriguing in view of the recent
theoretical proposal that there may be a ``superconducting thermal 
insulator" phase inside a superconductor \cite{Vishveshwara}.
To address this issue, it is desirable to measure the thermal
conductivity at very low temperatures in a series of
La$_{2-x}$Sr$_x$CuO$_4$ (LSCO) single crystals (where the behavior
of the low-temperature normal state is known) and trace the behavior
of the electronic thermal conductivity as $x$ is varied.

In this Letter, we report our measurements of the low-temperature
thermal conductivity in a series of LSCO single crystals
with $x = 0 - 0.22$; these $x$ values covers the whole range of 
the parent Mott insulator, the underdoped superconductor,
and the overdoped superconductor.
We find that the extended QPs appear to exist in the
zero-temperature limit throughout the superconducting phase
and that the transition to the ``thermally insulating" state takes
place concomitantly with the superconductor-to-insulator
transition at $x$ = 0.05.
Therefore, the low-energy physics of the QPs in the superconducting 
state and that of the charge carriers in the normal state under 60 T 
appear to be fundamentally different in the underdoped region. 
Moreover, the evolution of the QP heat transport in the superconducting 
state clearly demonstrates the shortcomings of the simple $d$-wave 
phenomenology. 

The single crystals of LSCO ($x = 0 - 0.22$) are grown by the
traveling-solvent floating-zone technique.
The underdoped (overdoped) crystals are annealed in reducing
(oxygenating) atmosphere to minimize deviation from the
stoichiometric oxygen content.
The crystals are cut into rectangular platelets with the typical
dimensions of $1.5 \times 0.5 \times 0.1$ mm$^3$, where the
$c$ axis is perpendicular to the platelets within an accuracy
of 1$^{\circ}$.
The optimally-doped sample ($x$=0.17) shows zero resistivity
at 40.5 K, and the most underdoped superconducting sample ($x$=0.06)
shows zero resistivity at $\sim$8 K.  The temperature dependences
of the in-plane resistivity $\rho_{ab}$ for most of the $x$ values
studied here are reported in Ref. \cite{mobility}.

The thermal conductivity $\kappa$ is measured using the conventional
steady-state ``one heater, two thermometer" technique.
A chip heater and two RuO$_2$ chip sensors are
attached to the sample with gold wires; on the sample, ohmic contacts
with the contact resistance of less than 0.5 $\Omega$ are made
with gold contact pads and the gold wires are attached to these
pads using silver epoxy.  To minimize heat leak,
superconducting NbTi wires with 15-$\mu$m diameter
are used as the leads of the chip sensors.
The lowest temperature of our thermal conductivity measurement
is typically 90--100 mK, below which it becomes uncertain whether
the true electron temperature is measured with our setup
assembled on a cold finger of a dilution refrigerator.
The temperature difference between the two thermometer contacts,
which are separated by $\sim$1 mm, is controlled to be typically
3\% of the sample temperature.
The error in $\kappa$ due to geometrical factors is less then 10\%.

Figure 1(a) shows the plots of $\kappa/T$ vs. $T^2$ for a series
of $x$ from 0 to 0.22.
If the electronic thermal conductivity $\kappa_{el}$ and the
phononic thermal conductivity $\kappa_{ph}$ behave as $a T$
and $b T^3$, respectively, like in ordinary Fermi liquids at
low temperatures, we can write $\kappa/T = a + b T^2$;
in this case, the zero-temperature intercept and the slope of a straight
line in the $\kappa/T$ vs. $T^2$ plot respectively give $a$ and $b$.
For cuprates, it has been demonstrated that $\kappa (T)$ obeys the
above temperature dependence for $T^2 < 0.02$ K$^2$ in typical 
crystalline samples \cite{Taillefer,Chiao,Behnia}, and the source of 
the $a T$ term in $\kappa$ of the superconducting cuprates has been 
discussed to be the Fermi-liquid-like impurity band 
created in the $d$-wave gap.

In Fig. 1(a), one can easily see the overall trend that $\kappa /T$
is shifted up with increasing $x$.
A natural extrapolation of the data for $T \rightarrow 0$ appears
to give a finite intercept for superconducting samples ($x > 0.05$),
and this intercept grows rather rapidly with $x$.
Therefore, the data in Fig. 1(a) already tell us that there is some finite
residual thermal conductivity $\kappa_{\rm res}$, which is of electronic
origin, in the superconducting samples and this $\kappa_{\rm res}$ tends to
grow as $x$ is increased above 0.05.

To quantify the above observation, we have tried to draw a
straight line that best fits the lowest-temperature part of
the $\kappa (T)/T$ data for each $x$.
The results are shown in Figs. 1(b)-1(m) for various $x$;
although for some of the $x$ values it is not obvious whether
the lowest-temperature data can be best described by a straight
line ($x$ = 0.08 and 0.14, for example), for many of the $x$ values
the data are actually well fitted with a straight line below
$T^2$ of $\sim$~0.02~K$^2$, which is in good correspondence with the
previous studies of other cuprates \cite{Taillefer,Chiao,Behnia}.
The zero intercept of the straight-line fit gives our best estimate
of $\kappa_{\rm res}/T$ for each $x$.
As is shown in Figs. 1(b) and 1(c), the data for the
non-superconducting samples extrapolates to essentially zero,
indicating that in these non-superconducting insulator samples
the low-temperature $\kappa$ is essentially phononic with negligible
electronic contribution; this is consistent with the observation in
insulating YBa$_2$Cu$_3$O$_{6.0}$ \cite{Taillefer}.
Note that $\kappa_{ph}$ becomes $\simeq b T^3$ when phonons
are predominantly scattered by the crystal surfaces \cite{Berman}, 
in which case $b$ is expressed as 
$\frac{1}{3}\beta \langle v_{ph} \rangle l_{ph}$, where, for LSCO, 
$\beta \simeq 3.9$ $\mu$J/cm$^3$ \cite{Fisher} is the 
phonon specific heat coefficient,
$v_{ph} \simeq 4 \times 10^5$ cm/s \cite{Dominec} is the averaged 
sound velocity, and $l_{ph}$ is the phonon mean-free path.
In perfect crystals $l_{ph}$ takes the maximum value $1.12\bar{w}$ 
with $\bar{w}$ the geometric mean width of the sample \cite{Thacher}. 
In our case, $\bar{w}$ is $170-300$ $\mu$m and the fits in 
Figs. 1(b)-1(m) yield $b$ of $2-6$ mW/cmK$^4$, giving 
$l_{ph}/1.12\bar{w}$ of $0.2 - 0.8$.
These ratios are a bit smaller than the ratios of $0.6-1.4$ for the 
Y-based cuprates \cite{Taillefer,Hussey} (which may be due to the 
roughness \cite{Thacher} of the polished surface or to the twins 
\cite{Berman}), but they are comparable to the 
Bi$_2$Sr$_2$CaCu$_2$O$_{8}$ (Bi-2212) case \cite{Nakamae} and 
are still in the reasonable range \cite{Berman}.

Figure 2 shows $\kappa_{\rm res}/T$ for all the samples measured 
as a function of $x$.
One can clearly see that $\kappa_{\rm res}/T$ starts to grow only
above $x$=0.05 with increasing $x$, and it is monotonically
increasing except for an anomalous dip at $x$=1/8.
The anomaly at $x$=1/8 is probably due to the charge ordering
\cite{Tranquada} which would tend to localize the carriers;
it is useful to note that a similar anomaly is observed in the
$x$ dependence of the penetration depth \cite{Panagopoulos1}.
The $\kappa_{\rm res}/T$ value at optimum doping ($x$=0.17) is
0.20 mW/cmK$^2$, which is comparable to the reported values 
for optimally-doped Y-123 \cite{Taillefer} and optimally-doped
Bi-2212 \cite{Chiao,Behnia}.

The most striking feature in Fig. 2 is probably the smooth
evolution of $\kappa_{\rm res}/T$ across optimum doping ($x \simeq 0.16$).
This is in sharp contrast to the behavior of the normal-state 
$\rho_{ab}$ under 60 T in the zero-temperature limit \cite{Ando,Boebinger},
which shows an insulator-to-metal crossover at optimum doping; namely,
$\rho_{ab}$ diverges as $T \rightarrow 0$ for $x < 0.16$, while
it stays small and finite for $x > 0.16$ \cite{Boebinger}.
This contrast can be more quantitatively illustrated by 
calculating the expected normal-state electronic thermal conductivity 
in the zero-temperature limit, $\kappa_{0}^{\rm normal}$, using 
the Wiedemann-Franz law $\kappa_{el}/T = L_{0}/\rho_{\rm res}$, 
where $L_{0} = (\pi^2/3)(k_B/e)^2$ is the Sommerfeld value of the
Lorentz number and $\rho_{\rm res}$ is the residual resistivity. 
(While the validity of this law is not clear in non-Fermi liquids, 
it should be satisfied in all Fermi liquids at low enough temperature 
\cite{Berman}.) 
Since $\rho_{ab}$ diverges for $x < 0.16$, $\kappa_{0}^{\rm normal}$ 
is zero in the underdoped region.  The data for $x$=0.17 and 0.22 
given in Ref. \cite{Boebinger} can be used to calculate 
$\kappa_{0}^{\rm normal}/T$ in the overdoped region, and the results are 
plotted in Fig. 2 with open squares; clearly, the normal-state 
$\rho_{ab}$ suggests $\kappa_{0}^{\rm normal}/T$ to grow rapidly 
above $x$=0.16, which is at odds with the behavior 
of $\kappa_{\rm res}/T$.
It is intriguing that $\kappa_{0}^{\rm normal}/T$ is larger than
$\kappa_{\rm res}/T$ in the overdoped region but their relation
switches in the underdoped region; 
in ``conventional" BCS superconductors with a $d$-wave gap, 
$\kappa_{el}$ in the normal state should always be much larger 
than that in the superconducting state, reflecting the difference 
in the numbers of available heat carriers.

To corroborate the above comparison, we further measured 
both the normal-state $\rho_{ab}(T)$ under 18 T
and $\kappa(T)$ in zero field in the identical sample for $x$=0.06.
The main panel of Fig. 3 shows the $\rho_{ab}(T)$ data and
the inset shows the comparison of $\kappa_{\rm res}/T$ with the 
``expected" normal-state electronic thermal conductivity 
$\kappa_{el}^{\rm normal}$ that would correspond to the $\rho_{ab}$ 
value; it is clear that $\kappa_{el}^{\rm normal}$ for 
$T \rightarrow 0$ is much smaller than $\kappa_{\rm res}/T$, 
which is very difficult to understand in the conventional picture 
of the superconducting condensate.

These comparisons suggest that either the scattering rate for 
$T \rightarrow 0$ in the superconducting state is much smaller than that 
in the normal state, or the Wiedemann-Franz law is strongly violated 
in the underdoped cuprates.  If the former is the case, it is a highly 
unusual situation, because any inelastic scattering would normally 
vanish as $T \rightarrow 0$ and the same elastic impurity scattering 
would dominate the transport in both the normal and the superconducting 
state; if, on the other hand, the latter is the case, our result 
is a yet-another strong proof of the non-Fermi-liquid nature of the 
normal state of the cuprates.  
In any case, the above observation highlights an unusual contrast 
between the low-energy physics of the superconducting state and that 
of the normal state in underdoped LSCO, and this contrast is useful 
in examining the nature of the strongly-correlated electrons in 
cuprates.  In fact, some of the existing theories offer intriguing 
possibilities to understand this unusual situation:
(a) The nature of the charge carriers may be fundamentally different 
between the normal and the superconducting states.
For example, it might be possible that the superconductivity 
occurs by avoiding a strong correlation effect that would otherwise
localize the holes \cite{Carlson,Hirsch}; in this case, the QPs
excited from the superconducting condensate may not be bound to the
correlation effect characteristic of the normal state.
(b) When there is a quantum critical point (QCP), inelastic scatterings 
can survive down to very low temperatures because of the critical 
fluctuations \cite{QCP}.  In this case, inelastic scattering may 
survive as $T \rightarrow 0$ in the normal state, which causes the 
total scattering rate in the normal state to be significantly larger.  
(c) It was argued \cite{Varma} that in a non-Fermi-liquid arbitrarily 
small concentration of impurities lead to a vanishing density of 
states at $E_F$ because of the strong electron correlations.  
In this case, the modification of the density of states may only be 
reflected in the heat carriers in the normal state.  
Of course, these three possibilities are just examples of the 
implication of the data, and our observation of the contrasting 
low-energy physics in the normal and the superconducting states 
would serve as a touchstone to test the theory of high-$T_c$ 
superconductivity.

In addition, the observed evolution of $\kappa_{\rm res}/T$ clearly 
demonstrates that the QP transport in the superconducting state is much 
more complicated than the simple $d$-wave phenomenology suggests.
The standard theory predicts \cite{Lee} 
that $\kappa_{\rm res}/T$ is proportional to the ratio $v_F/v_2$ 
($v_F$ and $v_2$ are the energy dispersion normal and tangential to the 
Fermi surface at the node) once the $d$-wave superconductivity is 
established, and it has been reported that the measured values of 
$\kappa_{\rm res}/T$ in Y-123 and Bi-2212 are in reasonable agreement 
with the theory at optimum doping \cite{Taillefer,Chiao,Behnia}; 
therefore, we would expect
that with increasing $x$ there is a discontinuous onset of
$\kappa_{\rm res}/T$ upon entering into the superconducting regime.
However, what we actually observe is 
a continuous and gradual increase in $\kappa_{\rm res}/T$ across 
$x$=0.05, which contradicts the theoretically expected behavior.
We note that the first indication for the breakdown of the simple 
phenomenology came from the measurements on Y-124 \cite{Hussey}, 
and the present result shows the breakdown of a different nature.  
Furthermore, a systematic photoemission study of Bi-2212 reported 
that $v_F/v_2$ decreases with carrier doping \cite{Mesot}, which 
suggests that $\kappa_{\rm res}/T$ as predicted by the theory should 
also {\it decrease} with increasing $x$.  
Hence, the results obtained here 
disagree with the theory in this respect as well, which suggests 
that either the agreement at optimum doping is accidental, or the 
doping dependence of the gap structure is not similar in different 
cuprates (which is improbable), or the standard theory breaks down 
in the underdoped regime.  In any case, the continuity at $x$=0.05 
appears to indicate that the 
QPs in the superconducting state are strongly influenced by the
localizing tendency in the normal state \cite{Ando}; namely, 
a part of the QPs somehow localize in the heavily-underdoped region 
and they cannot participate in the extended QP state.  
The mechanism of this QP localization in the superconducting state is 
not clear at this stage; however, the charge stripe ordering should
at least partly be responsible, because $\kappa_{\rm res}/T$ is
clearly suppressed at the 1/8 doping.  Other mechanisms such as
quantum interference effect \cite{Senthil1998}
might also be responsible.

In summary,
we find that the simple $d$-wave phenomenology 
for the QP transport in the superconducting state is not sufficient
to explain the observed evolution of $\kappa_{\rm res}/T$ with hole
doping $x$; specifically, the smooth onset of $\kappa_{\rm res}/T$
across $x$=0.05 suggests some additional mechanism that causes the
QP states to localize.
Moreover, we demonstrate that the low-energy physics shows a strong 
contrast between the normal and the superconducting states, 
which bears intriguing implications on the possible mechanism of 
superconductivity.

We greatly thank L. Taillefer and N. E. Hussey
for useful experimental information.
We also thank P. J. Hirschfeld and A. N. Lavrov for
helpful discussions and K. Segawa, T. Suzuki, and Y. Kurita
for technical assistance. X. F. S. acknowledges support from JISTEC.

%
%
%

\newpage
\widetext

\begin{figure}
\epsfxsize=16cm
\centerline{\epsffile{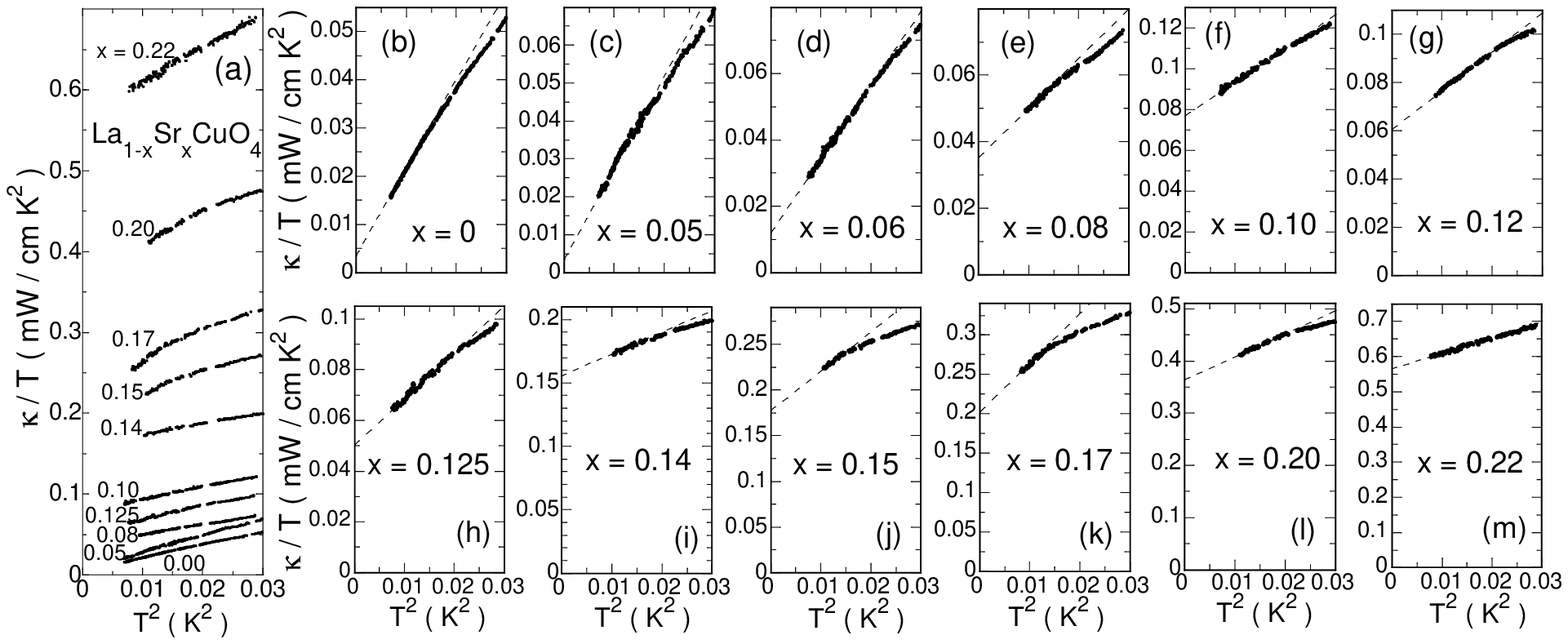}}
\vspace{0.3cm}
\caption{$\kappa/T$ of La$_{2-x}$Sr$_x$CuO$_4$ single crystals 
at low temperatures plotted vs. $T^2$. 
(a) Cumulative plot for selected $x$ values from 0 to 0.22. 
(b)-(m) Individual plots of $\kappa/T$ vs. $T^2$ for various $x$; 
dashed lines are linear fits to the low-$T$ part of the data.
Zero-temperature intercept of the dashed line gives the 
estimate of $\kappa_{\rm res}/T$.}
\label{fig1}
\end{figure}

\begin{figure}
\epsfxsize=7cm
\centerline{\epsffile{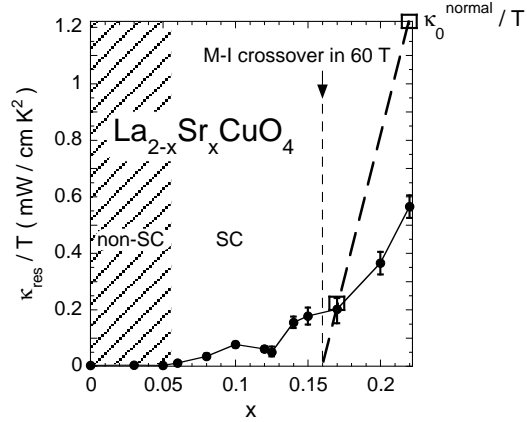}}
\vspace{0.3cm}
\caption{$\kappa_{\rm res}/T$ of LSCO as a function of $x$ 
(filled circles).  The expected normal-state electronic 
thermal conductivity for $T \rightarrow 0$, 
$\kappa_0^{\rm normal}$, is estimated from the 60-T data 
\protect\cite{Boebinger} and is shown for comparison (open squares).
Solid and dashed lines are guides to the eyes.}
\label{fig2}
\end{figure}

\begin{figure}
\epsfxsize=6cm
\centerline{\epsffile{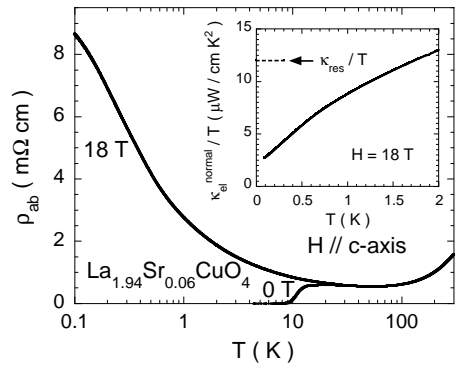}}
\vspace{0.3cm}
\caption{$T$-dependence of $\rho_{ab}$ in 0 and 18 T for $x$=0.06.
The same single crystal sample is used for the $\rho_{ab}(T)$ 
and $\kappa(T)$ [Fig. 1(d)] measurements.
The magnetic field is applied parallel to the $c$ axis.
Inset: The expected $\kappa_{el}^{\rm normal}(T)/T$ for $x$ = 0.06 
calculated from the $\rho_{ab}(T)$ data in 18 T, plotted together 
with $\kappa_{\rm res}/T$ estimated in Fig. 1(d).}
\label{fig3}
\end{figure}


\begin{references}
\vspace{30cm}


\bibitem{Volovik}
G. E. Volovik, JETP Lett. {\bf 58}, 469 (1993).

\bibitem{Lee}
A. C. Durst and P. A. Lee, Phys. Rev. B {\bf 62}, 1270 (2000), and 
references therein.

\bibitem{Moler}
K. A. Moler {\it et al.},
Phys. Rev. Lett. {\bf 73}, 2744 (1994).

\bibitem{Taillefer}
L. Taillefer, B. Lussier, R. Gagnon, K. Behnia, H. Aubin,
Phys. Rev. Lett. {\bf 79}, 483 (1997).

\bibitem{Chiao}
M. Chiao {\it et al.},
Phys. Rev. B {\bf 62}, 3554 (2000).

\bibitem{AndoBi2201}
Y. Ando {\it et al.}, Phys. Rev. Lett. {\bf 77}, 2065 (1996), and 
references therein.

\bibitem{Senthil1999}
T. Senthil and M. P. A. Fisher,
Phys. Rev. B. {\bf 60}, 6893 (1999).

\bibitem{Atkinson}
W. A. Atkinson, P. J. Hirschfeld, and A. H. MacDonald,
Phys. Rev. Lett. {\bf 85}, 3922 (2000).

\bibitem{Senthil1998}
T. Senthil, M. P. A. Fisher, L. Balents, and C. Nayak,
Phys. Rev. Lett. {\bf 81}, 4704 (1998).

\bibitem{Hussey}
N. E. Hussey {\it et al.},
Phys. Rev. Lett. {\bf 85}, 4140 (2000).

\bibitem{Ando}
Y. Ando, G. S. Boebinger, A. Passner, T. Kimura,
and K. Kishio,
Phys. Rev. Lett. {\bf 75}, 4662 (1995).

\bibitem{Boebinger}
G. S. Boebinger {\it et al.},
Phys. Rev. Lett. {\bf 77}, 5417 (1996).

\bibitem{Fournier}
P. Fournier {\it et al.},
Phys. Rev. Lett. {\bf 81}, 4720 (1998).

\bibitem{Ono}
S. Ono {\it et al.},
Phys. Rev. Lett. {\bf 85}, 638 (2000).

\bibitem{Vishveshwara}
S. Vishveshwara, T. Senthil, and M. P. A. Fisher,
Phys. Rev. B {\bf 61}, 6966 (2000).

\bibitem{mobility}
Y. Ando, A. N. Lavrov, S. Komiya, K. Segawa, and X. F. Sun,
Phys. Rev. Lett. {\bf 87}, 017001 (2001).

\bibitem{Behnia}
K. Behnia {\it et al.}, J. Low Temp. Phys. {\bf 117}, 1089 (1999).

\bibitem{Berman}
R. Berman, {\it Thermal Conduction in Solids}
(Oxford University Press, Oxford, 1976).

\bibitem{Fisher}
R. A. Fisher {\it et al.}, Phys. Rev. B {\bf 61}, 1473 (2000).

\bibitem{Dominec}
J. Dominec, J. Supercond. {\bf 3}, 337 (1990).

\bibitem{Thacher}
P. D. Thacher, Phys. Rev. {\bf 156}, 975 (1967).

\bibitem{Nakamae}
S. Nakamae {\it et al.}, Phys. Rev. B {\bf 63}, 184509 (2001).

\bibitem{Tranquada}
J. M. Tranquada, B. J. Sternlieb, J. D. Axe, Y. Nakamura,
and S. Uchida, Nature (London) {\bf 375}, 561 (1995).

\bibitem{Panagopoulos1}
C. Panagopoulos, B. D. Rainford, J. R. Cooper, C. A. Scott,
and T. Xiang, Physica C {\bf 341-348}, 843 (2000).

\bibitem{Carlson}
E. W. Carlson, D. Orgad, S. A. Kivelson, and V. J. Emery,
Phys. Rev. B {\bf 62}, 3422 (2000).

\bibitem{Hirsch}
J. E. Hirsch, Phys. Rev. B {\bf 62}, 14487 (2000).

\bibitem{QCP}
C. Castellani, C. Di Castro, and M. Grilli, 
Z. Phys. B {\bf 103}, 137 (1997); 
F. Onufrieva and P. Pfeuty, Phys. Rev. B {\bf 61}, 799 (2000).

\bibitem{Varma}
C. M. Varma, Phys. Rev. Lett. {\bf 79}, 1535 (1997).

\bibitem{Mesot}
J. Mesot {\it et al.},
Phys. Rev. Lett. {\bf 83}, 840 (1999).

\end{references}
\end{document}